\documentclass[aps,prapplied,showpacs,showkeys,twocolumn,floatfix]{revtex4-2}
\usepackage{longtable}
\usepackage{morefloats}
\usepackage[dvips]{graphicx}
\usepackage[dvipsnames,usenames]{xcolor}
\usepackage[most]{tcolorbox}
\usepackage{epsfig,graphicx,amsfonts,amsbsy}
\usepackage{amsmath,amsfonts,amsthm,amssymb}
\usepackage{appendix}
\usepackage{makeidx}
\usepackage{url}
\usepackage{verbatim}
\usepackage[bookmarksnumbered,pdfpagelabels=true,plainpages=false,colorlinks=true,linkcolor=blue,citecolor=blue,urlcolor=blue]{hyperref}
\usepackage[rightcaption]{sidecap}
\usepackage{array}
\usepackage{booktabs}
\usepackage{multirow}
\usepackage{floatrow}
\usepackage{tabularx}
\usepackage{cancel,soul,ulem}
\usepackage[customcolors]{hf-tikz}

\begin{document}

\title{Spin diffusion length associated to out-of-plane resistivity of Pt thin films in spin pumping experiments}
\author{C. Gonzalez-Fuentes}
\affiliation{Departamento de F\'{i}sica, Universidad T\'{e}cnica Federico
Santa Mar\'{i}a, Avenida Espa\~{n}a 1680, 2390123 Valpara\'{i}so, Chile}
\author{R. K. Dumas}
\affiliation{Quantum Design Inc., 10307 Pacific Center Court, San Diego, California, 92121 USA}
\author{B. Bozzo}
\affiliation{Instituto de Ciencia de Materiales de Barcelona-Consejo Superior de Investigaciones Cient\'{i}ficas,Bellaterra,Spain}
\author{A. Pomar}
\affiliation{Instituto de Ciencia de Materiales de Barcelona-Consejo Superior de Investigaciones Cient\'{i}ficas,Bellaterra,Spain}
\author{R. Henriquez}
\affiliation{Departamento de F\'{i}sica, Universidad T\'{e}cnica Federico
Santa Mar\'{i}a, Avenida Espa\~{n}a 1680, 2390123 Valpara\'{i}so, Chile}
\author{C. Garc\'ia}
\affiliation{Departamento de F\'{i}sica, Universidad T\'{e}cnica Federico
Santa Mar\'{i}a, Avenida Espa\~{n}a 1680, 2390123 Valpara\'{i}so, Chile}
\email{claudio.gonzalez@postgrado.usm.cl}
\date{\today }

\begin{abstract}
We present a broadband ferromagnetic resonance study of the Gilbert damping enhancement ($\Delta \alpha$) due to spin pumping  in NiFe/Pt bilayers.  The bilayers, which have negligible interfacial spin memory loss, are studied as a function of the Pt layer thickness ($t_{\text{Pt}}$) and temperature (100-293 K). Within the framework of diffusive spin pumping theory, we demonstrate that Dyakonov-Perel (DP) or Elliot-Yaffet (EY) spin relaxation mechanisms acting alone are incompatible with our observations. In contrast,  if we consider that the relation between spin relaxation characteristic time ($\tau_{\text{s}}$) and momentum relaxation characteristic time ($\tau_{\text{p}}$) is determined by a superposition of DP and EY mechanisms, the qualitative and quantitative agreement with experimental results is excellent.  
Remarkably, we found that $\tau_{\text{p}}$ must be determined by the out-of-plane electrical resistivity ($\rho$) of the Pt film and hence its spin diffusion length ($\lambda_{\text{Pt}}$) is independent of  $t_{\text{Pt}}$.
Our work settles the controversy regarding the $t_{\text{Pt}}$ dependence of $\lambda_{\text{Pt}}$ by demonstrating its fundamental connection with $\rho$ considered along the same direction of spin current flow.
\end{abstract}
\maketitle
\section{Introduction}

Thin film bilayers comprised of a high spin-orbit coupling (SOC) metal and metal–ferromagnet (NM/FM) thin films are central to some of the most interesting phenomena in contemporary spintronics, such as the spin pumping \cite{Tserkovnyak02,Tserkovnyak05}, the spin Hall effect (SHE) \cite{Saitoh06} and current induced spin-orbit torques  (SOTs) \cite{Manchon19}.

A key, and sometimes controversial, issue in current spintronics research is the correct quantification of  the spin diffusion length ($\lambda_{\text{s}}$) in the NM, \textit{i.e.} the characteristic length over which the spin current dissipates. For example, Pt, by far the most extensively studied material for generation and detection of spin currents, $\lambda_{\text{s}}$ exhibits a large dispersion of values in the literature \cite{Vlaminck13,Liu14,Lihui13,Kondou12,Azevedo11,Ando11}.
Moreover, the spin-flip scattering mechanisms that determine $\lambda_{\text{s}}$ are not completely understood. It has been generally assumed that the Elliot-Yaffet (EY) mechanism is the dominant mechanism in Pt, however, recent results have shown that both the  Dyakonov-Perel (DP) and EY spin relaxation mechanisms coexist at low temperatures \cite{Freeman18}.

The origin of the discrepancies in reported values of $\lambda_{\text{s}}$ may be related to \textit{how} has been measured.  For example, estimation of $\lambda_{\text{s}}$ has relied primarily on the measurement of inverse spin Hall effect (ISHE) voltage ($V_{\text{ISHE}}$) at FM/NM bilayers with variable NM thickness ($t_{\text{NM}}$) \cite{Mosendz10,Liu11,Feng12,Kondou12,Nakayama12,Gomez14,Ando11,Vlaminck13,Azevedo11}, or conversely, the detection of FM layer magnetization perturbations due to the SHE generated spin currents in the NM layer \cite{Hao15,Nguyen16,Du20}.
Such {\it{electrical detection methods}} always rely on the spin-to-charge current conversion factor, the so-called spin Hall angle ($\theta_{\text{SH}}$), to calculate $\lambda_{\text{s}}$. 
As example, in FM/NM bilayers, $V_{\text{ISHE}}$ has the form \cite{Ando11}:
\hfsetfillcolor{yellow!90}
\hfsetbordercolor{white}
\begin{equation}
V_{\text{ISHE}}\propto \theta_{\text{SH}}\lambda_{\text{s}},
\label{prop}
\end{equation}
which is valid for $t_{\text{NM}} \gg \lambda_{\text{s}}$. In Eq. (1), both $\theta_{\text{SH}}$ and $\lambda_{\text{s}}$ have a dependence on the resistivity of the NM layer ($\rho$).
On one hand, there is a  proportionality between $\theta_{\text{SH}}$ and $\rho$  which is characteristic of the intrinsic Berry-phase-related mechanism of SHE, currently accepted to be dominant in Pt \mbox{\cite{Guo08,Sagasta16,Isasa15,Morota16,Nguyen16,Zhu19}}. On the other hand, the connection of $\lambda_{\text{s}}$ with $\rho$ arises from its proportionality to the square root of characteristic momentum relaxation time ($\tau_{\text{p}}$), hence $\lambda_{\text{p}}\propto \rho^{-1/2}$ \cite{Nguyen16,Roy17}. 

At this point we emphasize a key detail that has not been addressed by previous works. If we assume $\theta_{\text{SH}}\propto \rho$, we expect $\rho$ corresponds to the \textit{in-plane} resistivity of the film ($\rho_{\parallel}$) given that the charge current flows in-plane. In a similar vain, if we have a spin current flowing out-of-plane of the FM layer, it is reasonable to expect that $\lambda_{\text{s}}$ will be then be determined by the \textit{out-of-plane} resistivity ($\rho_{\perp}$) of NM. With this reasoning, an expression of the form $V_{\text{ISHE}}\propto \theta_{\text{SH}}(\rho_{\parallel}) \lambda_{\text{s}} (\rho_{\perp})$ should be employed to model the experimental data.  Curiously $\rho_{\parallel}$ has been employed not only to account for the $t_{\text{NM}}$ dependence of $\theta_{\text{SH}}$ but also for $\lambda_{\text{s}}$ dependence\, assuming implicitly that $V_{\text{ISHE}}\propto \theta_{\text{SH}}(\rho_{\parallel}) \lambda_{\text{s}} (\rho_{\parallel})$. We should also consider that whereas $\rho_{\perp}$ is similar to the bulk value and therefore independent of $t_{\text{NM}}$, $\rho_{\parallel}$ is remarkably higher and has exhibits a strong dependence on $t_{\text{NM}}$ due reduced dimensionality.

An alternative method to extract $\lambda_{\text{s}}$ is with an experimental method that relies solely on the spin pumping effect and is therefore unaffected by the in-plane generated charge currents and the value of $\theta_{\text{SH}}$. Analyzing the Gilbert damping ($\alpha$) enhancement due to the NM layer in ferromagnetic resonance experiments accomplishes this goal.  If we fix the thickness of the FM layer and vary $t_{\text{NM}}$, the extracted $\alpha$ vs $t_{\text{NM}}$ will follow a characteristic exponential saturation curve \cite{Tserkovnyak02,Tserkovnyak05,Foros05}. Although this is a well known technique, the precise quantification of $\lambda_{\text{s}}$ requires a series of bilayers with a small to negligible interfacial spin memory loss (SML) \cite{Rojas-Sanchez14,Liu14} which is uncommon. Therefore, this requirement has difficult reliable quantification of $\lambda_{\text{s}}$, because if SML is included in the analysis at least two additional adjustable parameters, are required, namely the spin mixing conductance ($g^{\uparrow\downarrow}$) and interfacial depolarization parameter ($\delta$).


In this work we aim to settle the
issue of the $t_{\text{NM}}$ dependence of $\lambda_{\text{s}}$ in Pt by employing variable temperature spin pumping experiments and analysing the Gilbert damping. The negligible SML of our samples provides us the opportunity to perform an analysis which integrates the $\rho$ vs T dependence of Pt into $\lambda_{\text{s}}$ and test the diffusive model of Tserkovnyak \textit{et al.} \cite{Tserkovnyak02,Tserkovnyak05}.
Our work provides strong evidence that $\rho_{\perp}$ determines $\lambda_{\text{s}}$ and hence it does not depend on $t_{\text{NM}}$. In addition, we show evidence of the low influence of temperature in $\tau_{\text{s}}$ of Pt, in concordance with recently reported results \mbox{\cite{Freeman18}}.

\section{Experimental details}
A series of NiFe(20)/Pt($t_{\text{Pt}}$) bilayers were deposited onto thermally oxidized SiO$_2$ substrates by DC magnetron sputtering, with $t_{\text{Pt}}$=1, 2, 4 and 7 (all thicknesses expressed in nm). 
The Ar sputtering gas pressure and power during deposition was 1 mTorr and 60 W, respectively.  To promote better uniformity, the substrates were rotated during growth.
A NiFe(20 nm)/Ag(2 nm) reference sample was also deposited under the same conditions. As the reported value of $\lambda_{\text{s}}$ for Ag is approximately 450 nm \cite{Isasa15a}, we can confidently neglect spin attenuation effects in this reference sample. The samples were finally annealed \textit{in-situ} at 500 K for 1 hour in vacuum. This procedure has been found to substantially improve the interfacial spin conductance, as evidenced by the onset of the characteristic $\Delta \alpha$ vs $t_{\text{NM}}$ curve as compared to to non-annealed samples. Further details of this annealing study will be addressed in a future work. 

Also, in order to check the good coverage of the NiFe by Pt overlayer in our samples, we have carried out Scanning Tunneling Microscopy Measurements of the $t_{\text{Pt}}$=7 nm sample (appendix \mbox{\ref{app2}}).

\begin{figure}[ht]
\includegraphics[width=8.5cm]{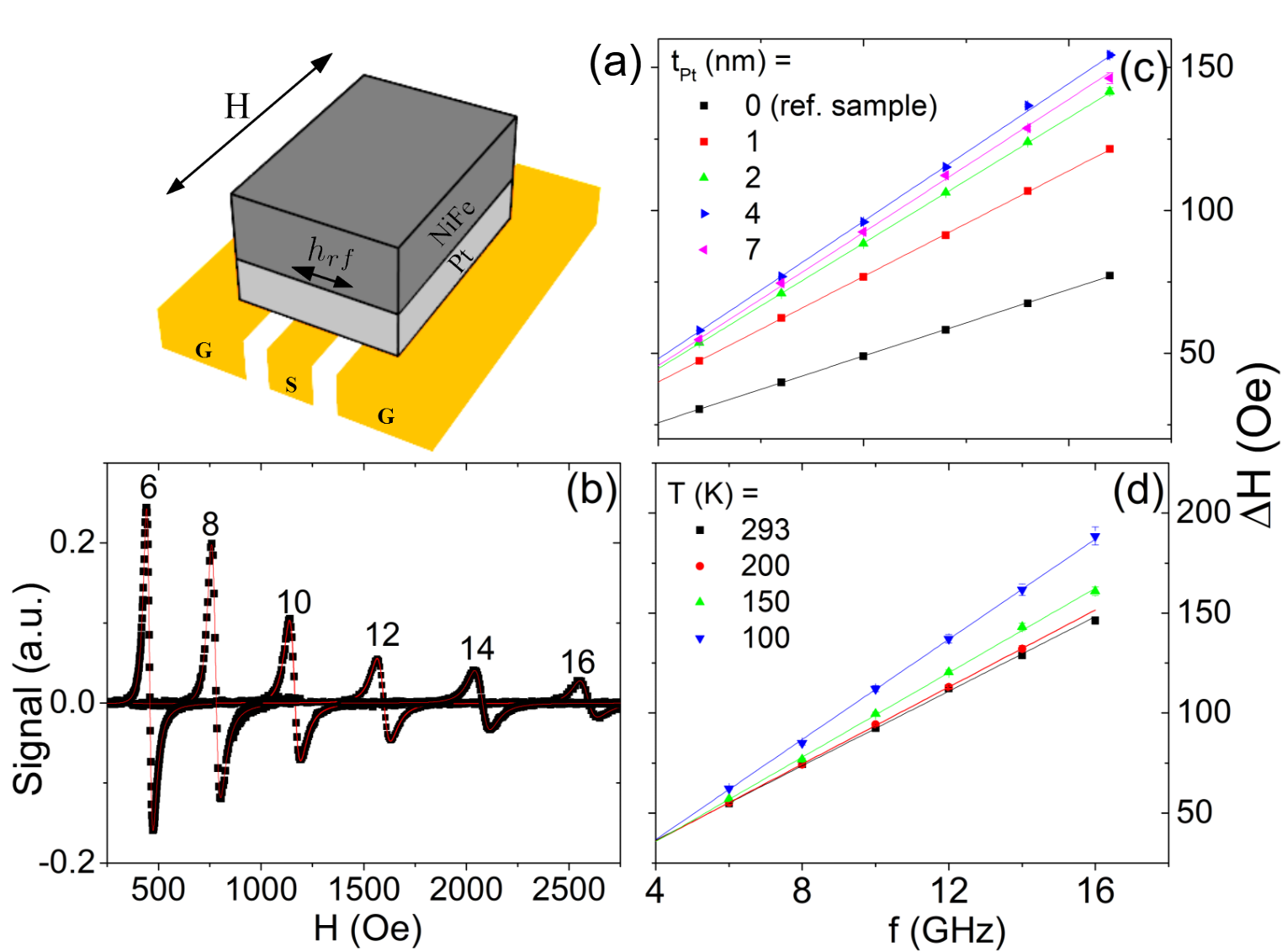}
\caption{(a) Schematic of the broadband measurement system, (b) FMR spectra of the NiFe(20 nm)/Pt(7 nm) film at 293 K for different values of $f$ in GHz, (c) FMR peak linewidths vs $f$ at room temperature for NiFe(20 nm)/Pt$(t_{\text{Pt}})$ for $t_{\text{Pt}}$ = 1,2,4 and 7 nm and Ag capped reference sample.
(d) $\Delta H$ vs $f$  for NiFe(20 nm)/Pt(7 nm) film at $T$ = 100, 150, 200 and 293 K. In (c) and (d) continuous lines are linear fittings to the experimental points.} 
\label{Fig1}
\end{figure}

The thin film bilayer samples were characterized by broadband (6-16 GHz) ferromagnetic resonance at 100, 150, 200 and 293 K.  The measurements are performed at a fixed frequency while sweeping the magnetic field through resonance.  Due to the field-modulated measurement scheme, the derivative of the microwave absorption is recorded. The obtained resonance spectra are then fitted to the sum of a symmetric and asymmetric Lorentzian function:
\hfsetfillcolor{yellow!90}
\hfsetbordercolor{white}
\begin{equation}
 P=\frac{d}{dH}\left[\frac{S(\Delta H)^2+A_{S}(H-H_{r})}{4(H-H_{r})^{2}+(\Delta H)^2}\right],
\label{Lorentzian}
\end{equation}
were $H_{r}$, $\Delta H$, $S$ and $A_{S}$ are the fitting parameters corresponding to the resonance field, the linewidth, the symmetric and the asymmetric components of the Lorentzian curve, respectively. The Gilbert damping was obtained by a linear fit of the frequency dependence of the extracted linewidth:
\begin{equation}
\Delta H=\Delta H_{0}+\frac{2h\alpha f}{g \mu_{\text{B}}\mu_{0}}
\end{equation}
where $\Delta H_{0}$, $\mu_{0}$, $\mu_{\text{B}}$, $h$, $g$ are the inhomogeneous broadening, vacuum permeability, Bohr magneton, Plank's constant and gyromagnetic factor ($g=2.11$ for NiFe \cite{Shaw11}), respectively. Additional details of data analysis protocol can be found in Ref. \cite{Gonzalez-Fuentes18}. Figure \ref{Fig1} shows representative graphs of these results. 

\begin{figure}[ht]
\includegraphics[width=8.5cm]{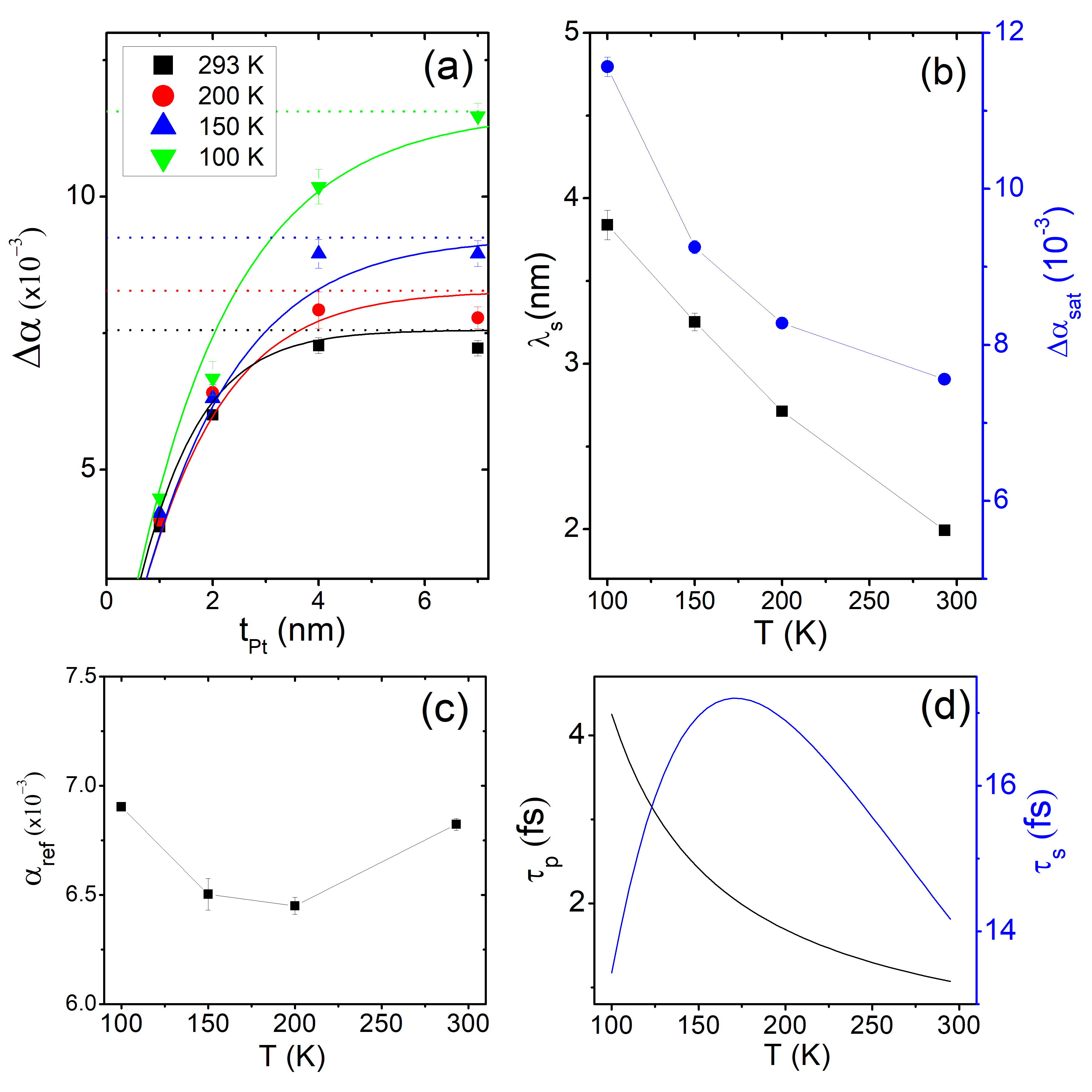}
\caption{(a) $t_{\text{Pt}}$-dependence of $\alpha $ for NiFe(20 nm)/Pt(t$_{\text{Pt}}$) bilayers at $T=$ 100, 150, 200, and 293 K. Symbols are the experimental data and continuous lines are fitting to the EY-DP model employed on this work. Dotted horizontal lines are at corresponding values of $\Delta \alpha_{\text{sat}}$. (b) $T$-dependence of $\lambda_{\text{s}}$  and $\Delta \alpha_{\text{sat}}$ obtained from the EY-DP model, (c) experimental $T$-dependence of $\alpha$ for Ag capped reference sample, (d) $T$-dependence of $\tau_{\text{p}}$, derived from Eq. \mbox{\eqref{taup}}, and $\tau_{\text{s}}$ derived from the EY-DP model.}.
\label{Fig2}
\end{figure}
\section{Results and discussion}
The NiFe/Pt bilayers show an increased $\alpha$ in comparison to the NiFe/Ag reference sample. This difference is due to the imbalance between the amount of angular momentum carried by the electron spins pumped into the NM, and the portion flowing back into the FM.
The additional damping contribution, $\Delta\alpha$, was quantified by simply subtracting the damping of the Ag reference sample ($\alpha_{0}$) from the damping of the NiFe/Pt bilayers, $\Delta\alpha=\alpha-\alpha_{0}$.
We modeled the $\Delta \alpha$ vs $t_{\text{Pt}}$ curves using the diffusive theory of spin pumping \cite{Tserkovnyak02,Tserkovnyak05}, in which the additional damping contribution ($\Delta \alpha$) of the NiFe/Pt bilayers is given by:
\begin{equation}
\Delta\alpha=\alpha-\alpha_{0}=\frac{g\mu_{B}g^{\uparrow \downarrow}_{\text{eff}}}{4\pi M_{\text{s}}t_{\text{F}}}\text{;   }g^{\uparrow \downarrow}_{\text{eff}}=\frac{g^{\uparrow\downarrow}}{1+g^{\uparrow\downarrow}\beta}
\label{da}
\end{equation}
where  $g$, $\mu_{\text{B}}$, $M_{\text{s}}$ and $t_{\text{F}}$ are the gyromagnetic  factor, Bohr magneton, saturation magnetization and thickness of the ferromagnet, respectively. $g^{\uparrow \downarrow}_{\text{eff}}$ is the \textit{effective} spin mixing conductance of the interface, defined in terms of the intrinsic spin mixing conductance $g^{\uparrow \downarrow}$ and the back-flow factor:
\begin{equation}
\beta=\frac{\tau_{\text{s}} \tanh^{-1}\left(L/\lambda_{\text{s}}\right)}{h \nu \lambda_{\text{s}}},
\label{beta}
\end{equation}
where $\nu$ is the single spin density of states of NM. Note, $\alpha_{0}$ does not show significant dependence on temperature, as shown in Fig. \ref{Fig2}. 

$\lambda_{\text{s}}$ can be expressed by the  relation:
\begin{equation}
\lambda_{\text{s}}=v_{F}\sqrt{\frac{\tau_{\text{s}}\tau_{\text{p}}}{3}}
\label{lambda}
\end{equation}
where $v_{F}$ is the Fermi speed, $\tau_{\text{s}}$ and $\tau_{\text{p}}$ are the characteristic spin and momentum relaxation times, respectively. 

As explained in the appendix \ref{app1}, SML \cite{Rojas-Sanchez14,Boone15,Liu14} can be neglected in our samples.
The saturation value of $\Delta \alpha$ is defined as $t_{\text{NM}}\rightarrow\infty$:
\begin{equation}
\Delta \alpha_{\text{sat}}=\frac{g\mu_{B}g^{\uparrow\downarrow}}{4\pi M_{\text{s}}t_{\text{F}}}\left(1+\frac{g^{\uparrow\downarrow}}{h \nu v_{F}}\times\sqrt{\frac{3\tau_{\text{s}}}{\tau_{\text{p}}}}\right)^{-1}.
\label{sat}
\end{equation}
From a physical point of view, the positive correlation of $\Delta\alpha_{\text{sat}}$  with $\tau_{\text{p}}$ arises from to the increment of the  diffusion coefficient $D=\lambda_{\text{s}}^{2}/\tau_{\text{s}}$ \mbox{\cite{Tserkovnyak02}} of the spin polarized electrons in the NM layer. This effect overcomes the negative correlation between $\Delta\alpha_{\text{sat}}$ and $\tau_{\text{p}}$ arising from the enhancement of $\lambda_{\text{s}}$ and the concurrent increase of spin current backflow. Overall, this translates into an increment of $\Delta \alpha_{\text{sat}}$ as $T$ goes down.

Regarding the correlation between $\tau_{\text{p}}$ and $\tau_{\text{s}}$, there have two mechanisms for spin-flip scattering have been proposed:  Elliot-Yaffet \cite{Elliot54,Yafet83} (EY) and Dyakonov-Perel \cite{Dyakonov71,Mower11} (DP). In the EY picture, each momentum scattering event has a probability of being a spin-flip event ($\tau_{\text{s}} \propto \tau_{\text{p}}$), whereas in the DP picture, the spin de-phasing occurs continuously so $\tau_{\text{s}} \propto \tau_{\text{p}}^{-1}$.

The temperature variation of $\Delta \alpha_{\text{sat}}$ provides valuable insight into the physics of the various scattering mechanisms in Pt, since $\tau_{\text{p}}\propto\rho^{-1}$ and $\rho$ increases in an approximately linear manner with temperature over the range of temperatures studied here.
If, for example, we assume the EY model alone is valid, according to Eq. \eqref{lambda}, $\lambda_{\text{s}}$ should decrease with increasing temperature, in concordance with previous predictions and observations \cite{Nguyen16,Liu15,Freeman18}. In this framework, according to Eq. \eqref{sat},  $\Delta\alpha_{\text{sat}}$ should be independent of temperature owing to the cancellation of $\tau_{\text{p}}$ in the right side of the equation. This clearly contradicts our experimental results: in Fig \ref{Fig2}{\color{blue}{(a)}}, a two-fold increase of $\Delta\alpha_{\text{sat}}$ in the temperature range from 100 K to 293 K is observed.
On the other side, if we assume the DP model is the exclusive spin-flip scattering mechanism, it leads to the unacceptable outcome that $\lambda_{\text{s}}$ is independent of temperature, in this case from the cancellation of $\tau_{\text{p}}$ in the right side of  Eq. \mbox{\eqref{lambda}}. This model, however, is concordant with our observation of increasing $\Delta\alpha_{\text{sat}}$ as $T$ decreases predicted by Eq. \eqref{sat}.

Note, these conclusions are still valid for models that also include SML at the interface \cite{Liu14,Rojas-Sanchez14,Tao18}, despite the fact that SML narrows the range of variation of $g^{\uparrow \downarrow}_{\text{eff}}$ vs $t_{\text{NM}}$.

Although it is generally accepted that EY should be the dominant mechanism of spin-flip scattering in crystal systems with cubic symmetry, the evidence is not conclusive. For example, recently evidence of dominant DP \cite{Boone15,Ryu16} and partial contributions of both DP and EY mechanisms \cite{Dai19,Long16,Ryu16} in poly-crystalline Pt films have been found.
Boone \textit{et al.} \cite{Boone15} performed an extensive analysis of spin pumping in NiFe/Pt and NiFe/Pd bilayers and found that their data is better modeled by a DP model. Villamor \textit{et. al.}  found significant deviations from EY mechanism in Cu \cite{Villamor13}. Very recently Freeman \textit{et al.} \cite{Freeman18} found evidence of DP mechanism in Pt at cryogenic temperatures. The authors interpreted their results as a near compensation of the DP and EY mechanisms for a broad range of temperatures, such that $\tau_{\text{s}}$ varies only a 20\% from 50 K to room temperature. Coincidentally, our results suggest that a mixed or intermediate model between DP and EY can account for our qualitative observations.

To test this hypothesis we modeled $\tau_{\text{p}}$ in Eq. \eqref{sat}, using the Sommerfield model:
\begin{equation}
\tau_{\text{p}}=3\left(q_{e}^{2}\nu v_{F}^{2}\rho\right)^{-1}.
\label{taup}
\end{equation}

Here, in contrast to other works \cite{Boone15,Freeman18}, we have chosen not utilize the Dr\"{u}de approximation since it assumes a spherical Fermi surface of Pt, which is an over simplification and inadequate approximation \cite{Dye78}.

We can narrow the possible range of variation of $\rho$ by using Eqs. \eqref{sat} and \eqref{taup} in a given range of variation of $\Delta\alpha_{\text{sat}}$ even without knowing the explicit value of $\rho(T)$. Defining $\rho_{100}$, $\rho_{293}$, $\Delta\alpha_{\text{sat,100}}$ and $\Delta\alpha_{\text{sat,293}}$ as the values of resistivity and $\Delta\alpha_{\text{sat}}$ at 100 K and 293 K respectively, we can derive the following inequality from Eq. \eqref{sat}:
\begin{equation}
\left(\frac{\Delta\alpha_{\text{sat},100}}{\Delta\alpha_{\text{sat},293}}\right)^{2}\leq\frac{\rho_{293}}{\rho_{100}}.
\label{max}
\end{equation}

Central to this work is the question of which resistivity, \textit{i.e.} in-plane $\rho_{\parallel}$ or out-of-plane $\rho^{\perp}$, determines $\lambda_{\text{s}}$.

Apart from the strong thickness dependence of $\rho_{\parallel}$, a key difference with respect to $\rho_{\perp}$ is the temperature dependence. The $T$-variation of $\rho_{\parallel}$ is weaker than $\rho_{\perp}$, which is mainly due to the strong contribution of the temperature independent scattering mechanisms such as electron-surface and electron grain-boundary, in comparison to the electron-phonon scattering that tend to be dominant in bulk films at temperatures higher than 50 K \cite{Mayadas70,Kastle04}.
In particular, for bulk Pt, the reported temperature ratio $\rho_{293}/\rho_{100}$ is approximately 3.6 \cite{Poker82,Arblaster15}, while for $\rho_{\parallel}$, is only up to 1.5 in 20 nm thick sputtered samples \cite{Sagasta16} Given that $T$-dependence of $\rho_{\parallel}$ decreases with decreasing $t_{\text{NM}}$ \mbox{\cite{Agustsson08,Henriquez19}}, $\rho_{293}/\rho_{100}$ must be even smaller in our films.
 We experimentally obtained a value of $(\Delta\alpha_{\text{sat},100}/\Delta\alpha_{\text{sat},293})^{2}\sim2.25$ which sets a lower limit to $\rho_{293}/\rho_{100}$. Consequently, the commonly accepted assumption that $\rho_{\parallel}$ determines $\lambda_{\text{s}}$ is hardly reconcilable with our experimental observations.

The assumption of a dominant EY mechanism in $\tau_{\text{s}}$ would reduce the possible range of variation of $\Delta \alpha_{\text{sat}}$ across temperature, owing to the $\tau_{\text{p}}/\tau_{\text{s}}$ ratio of the backflow factor in Eq. \eqref{sat}. Consequently, we would need to increase the lower possible limit of $\rho_{293}/\rho_{100}$ in order to compensate the former effect to be compatible with our experimental results. Similarly, if we include SML in our analysis, given that this effect always reduce the influence of the spin-current backflow over the value of $\Delta \alpha_{\text{sat}}$ \cite{Rojas-Sanchez14}, which is ultimately the origin of the $\alpha_{\text{sat}}$ variation across temperature. In summary, we conclude that $\lambda_{\text{s}}$ has to be determined by $\rho_{\perp}$ instead of $\rho_{\parallel}$. Furthermore, in our analysis, a greater contribution of the EY mechanism or SML would strengthen this conclusion.

Our experimental data points are then fit assuming $\rho_{\perp}$ in Eq. \eqref{taup} as the total resistivity of our system, equivalent to the bulk resistivity of Pt \cite{Poker82}.
The value of $M_{\text{s}}=0.94\pm0.05\text{ T}$ was extracted from the FMR measurements. $\nu=\text{1.1}\times10^{48}$ and $v_{F}=8.8\times 10^{5}\text{m/s}$ were taken from Refs. \cite{Jiao13} and \cite{Ketterson70}, respectively. We modeled the spin relaxation in Pt as a superposition of EY and DP mechanisms \mbox{\cite{Freeman18}} (EY-DP model): $\tau_{\text{s}}^{-1}=\tau_{\text{EY}}^{-1}+\tau_{\text{DP}}^{-1}$ where $\tau_{EY}=c_{\text{EY}}\tau_{\text{p}}$ and $\tau_{\text{DP}}=c_{\text{DP}}\tau_{\text{p}}^{-1}$ are the characteristic spin relaxation times associated to EY and DP mechanisms respectively. The coefficients $c_{\text{EY}}$ and $c_{\text{DP}}$ are the parameters to model their effective strengths. They, in combination with $g_{\uparrow\downarrow}$ were the set of adjustable parameters for the fitting of the all the experimental data. The results were $g_{\uparrow\downarrow}=(2.26 \pm 0.06)\times10^{20}\text{ m}^{-2}$, $c_{\text{EY}}=16.8 \pm 0.3$ and $c_{\text{DP}}=(7.03 \pm 0.06)\text{ s}^{2}$, with $R^{2}=0.9976$ for the fit.
The fitted curves are shown in Fig. \ref{Fig2}{\color{blue}{(a)}} and the corresponding values of $\lambda_{\text{Pt}}$ and $\Delta\alpha_{\text{sat}}$ are found in Fig. \ref{Fig2}{\color{blue}{(b)}}, whereas the expected temperature dependence of $\tau_{p}$ according to Eq. \eqref{taup} and $\tau_{\text{s}}$ according to EY-DP model is shown in Fig. \ref{Fig2}{\color{blue}{(d)}}. The excellent agreement of our predictions with the experimental data is remarkable given the simplicity of our assumptions. The connection between $\rho_{\perp}$ and bulk resistivity is also in concordance with structural analysis of the $t_{\text{Pt}}=$7 nm sample as shown in appendix \mbox{\ref{app2}.}

Our results are robust since spin pumping experiments based on the Gilbert damping are not affected by the value of $\theta_{\text{SH}}$ or $\rho_{\parallel}$.  Consequently, many of the potential problems that can influence measurements based on electrical detection are simply not present here. This offers a new perspective to the discussion that appeared recently regarding the thickness dependence of $\rho_{\parallel}$ and its influence on $\lambda_{\text{s}}$. In particular, it has been proposed that the neglecting this dependence may promote the large spread of $\lambda_{\text{s}}$ values present in the literature \cite{Nguyen16,Roy17,Swindells19}. Consequently, many recent works explicitly consider a thickness dependent $\lambda_{\text{s}}$ \cite{Roy17,Montoya16,Nguyen16,Zhu18,Swindells19,Ou16j}.

We believe that this mistake arises in great part because, whereas $\lambda_{\text{s}}$ is determined by $\rho_{\perp}$ and $\theta_{\text{SH}}$ is determined by $\rho_{\parallel}$, both $\rho_{\perp}$ and $\rho_{\parallel}$ have been treated {\it{indistinctly}}. 


 A further confirmation of our hypothesis arises when we review the published values of $\lambda_{\text{Pt}}$ in works that obtain it by from the $\alpha$($t_{\text{Pt}}$) curve and in bilayers with small SML. The dispersion of $\lambda_{\text{s}}$ values in works that satisfy these criteria  \cite{Caminale16,Swindells19,Huo17,Belmeguenai18}, is from 1.6 to 1.8 nm at room temperature (including the present work). Analogously, the published values of $\lambda_{\text{s}}$ of Pd applying the same selection criteria ranges from 6 to 9 nm \cite{Foros05,Kumar17,Caminale16}. Generally, the values of  $\lambda_{\text{s}}$ obtained by spin pumping experiments are more consistent for a given material as compared to those found by electrical detection methods.
An additional check of the consistency of the $t_{\text{NM}}$-independent hypothesis can be found in the recent work of Swindells et al. \mbox{\cite{Swindells19}}, on which the authors reported values of $\lambda_{\text{Pt}}$ obtained by spin  pumping enhanced $\alpha$,  in a series of bilayers consisting on Pt in combination with different FM materials. The work shows a direct comparison of the fitted value of $\lambda_{\text{Pt}}$ with $t_{\text{NM}}$-dependent and $t_{\text{NM}}$-independent expressions. In the second case the values of $\lambda_{\text{Pt}}$ were fairly more similar among the systems studied, rounding 1.6 nm in all cases. In comparison $t_{\text{NM}}$-dependent fitting gave a variation between 6.6 to 9.5 nm for bulk $\lambda_{\text{Pt}}$.

Our findings also make us discard an hypothesis of spin relaxation in Pt based on DP or EY mechanisms acting alone. A similar finding was reported in Ref. \mbox{\cite{Freeman18}}, and as is our work, they find that the $T$ variation of $\tau_{\text{s}}$ is much smaller than $\tau_{\text{p}}$. In that work the existence of DP mechanism in Pt was supported also  with evidence from  magnetoresistance measurements. However, a theoretical explanation of how this could exist in the centrosymmetric fcc structure of Pt is not present. Another option may be an intermediate spin relaxation regime between EY and DP rather than a mixed one. A recent work \mbox{\cite{Boross13}} has proposed that the characteristic relations $\tau_{\text{s}}\propto\tau_{\text{p}}$ from EY and $\tau_{\text{s}}\propto\tau_{\text{p}}^{-1}$ from DP spin relaxation mechanisms, are the limit cases of a broad spectrum of spin relaxations regimes. In this line, the unexpectedly small \textit{effective} correlation between $\tau_{\text{s}}$ and $\tau_{\text{p}}$ observed in this work and in Ref. \mbox{\cite{Freeman18}}, would mean simply that an intermediate relation ($\tau_{\text{s}}\propto\tau_{\text{p}}^{0}$)  is, at least, more appropriate to describe the actual spin relaxation in Pt. We expect that our results will motivate further theoretical investigations in this respect.

\section{Conclusions}
In summary, we have shown via temperature dependent spin pumping experiments that, in the framework of the diffusion theory, the spin diffusion length is determined by the out-of-plane resistivity of the NM layer and hence is not dependent on its thickness.
Our results also support the recent findings showing the impossibility
to explain $\tau_{\text{s}}$ of Pt exclusively by EY or DP mechanisms, suggesting a mixed or intermediate spin relaxation regime between them.

We believe that the apparent controversy regarding the different values of $\lambda_{\text{s}}$ found in SHE experiments must be due to the $t_{\text{NM}}$ dependence of $\theta_{\text{SH}}$ rather than $t_{\text{NM}}$ dependence of $\lambda_{\text{s}}$. The first is proportional to $\rho_{\parallel}$ and hence very variable with fabrication conditions, whereas the second depends on $\rho_{\perp}$ which is close to bulk resistivity, independently of $t_{\text{NM}}$.

\begin{appendices}
\appendix
\section{Negligible Spin Memory Loss in NiFe/Pt bilayers}
\label{app1}

We confirmed the negligible interface Spin Memory Loss (SML) in our samples comparing the extrapolated value of Gilbert's damping $\alpha$ at $t_{Pt}\rightarrow 0$, namely $\alpha_{\text{tPt}\rightarrow 0}$, and the measured value of $\alpha$ of the reference sample ($\alpha_{\text{ref}}$). It should be noted that the extra-damping generated by the normal metal: $\Delta \alpha(t_{\text{NM}})$ is non-zero at $t_{\text{NM}}=0$  in all the models which consider finite SM \cite{Rojas-Sanchez14,Liu14,Boone15,Chen15,Chen15b,Tao18}, and its value is often comparable to $\Delta \alpha(t_{\text{NM}}\rightarrow \infty)$ \cite{Rojas-Sanchez14,Azzawi16}.

\begin{figure}[h]
\includegraphics[width=8.5cm]{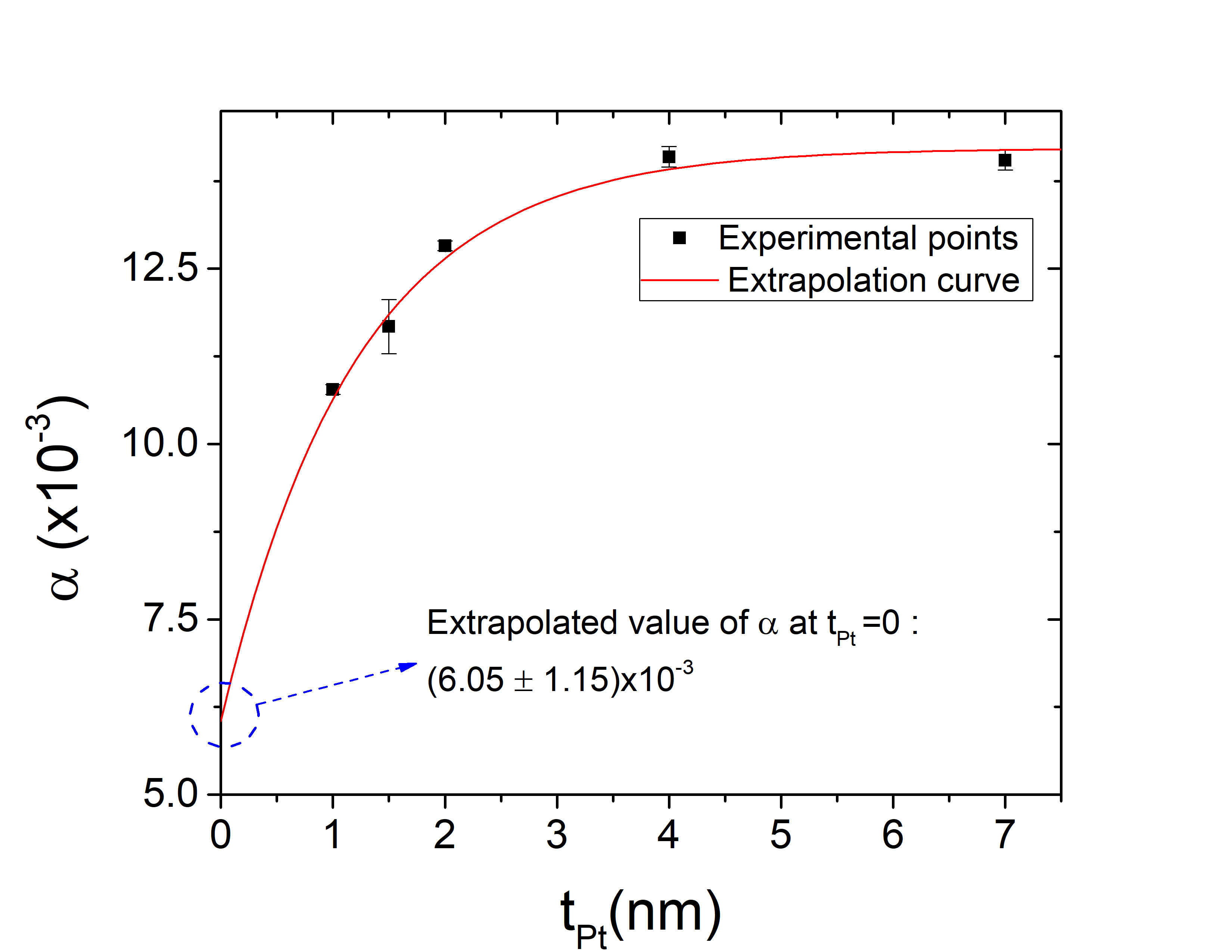}
\caption{$\alpha$ vs $t_{\text{Pt}}$ curve at room temperature (293 K) for the NiFe(20nm)/Pt($t_{Pt}$) series of samples (black squares) and SML extrapolation curve (red continuous line).}
\label{SML}
\end{figure}

To get $\alpha_{\text{tPt}\rightarrow 0}$ we employed the following extrapolation function to fit $\alpha$ vs $t_{\text{Pt}}$ curve at room temperature (Fig. \ref{SML}) :
\begin{equation}
\alpha\left(t_{\text{Pt}}\right)=\alpha_{\text{tPt}\rightarrow 0}+\alpha_{\text{s}}\exp\left(\frac{-t_{\text{Pt}}}{t_{\text{sat}}}\right),
\label{extrapol}
\end{equation}
where $\alpha_{t\rightarrow 0}$, $t_{\text{sat}}$, $\alpha_{\text{s}}$ are the fitting parameters.  
The obtained value of $\alpha_{\text{tPt}\rightarrow 0}$ was $\left(6.05\pm1.5\right)\times10^{-3}$ which is below the value of the reference sample: $\alpha_{\text{ref}}=\left(6.8\pm0.1\right)\times10^{-3}$, but inside the fitting uncertainty margin. 

We emphasize that we do not give any physical meaning to $\alpha_{\text{s}}$ nor $t_{\text{sat}}$ in \eqref{extrapol}, as it serves only for the extrapolation purpose.
In this sense, our method is analogous to other extrapolation methods employed for quantifying interfacial magnetic effects such as proximity effect \cite{Thorarinsdottir19} or the interfacial perpendicular anisotropy energy density \cite{Ikeda10}. 

\section{Scanig Tunneling Microscopy Study}
\label{app2}

\begin{figure}[h]
\includegraphics[width=8.0cm]{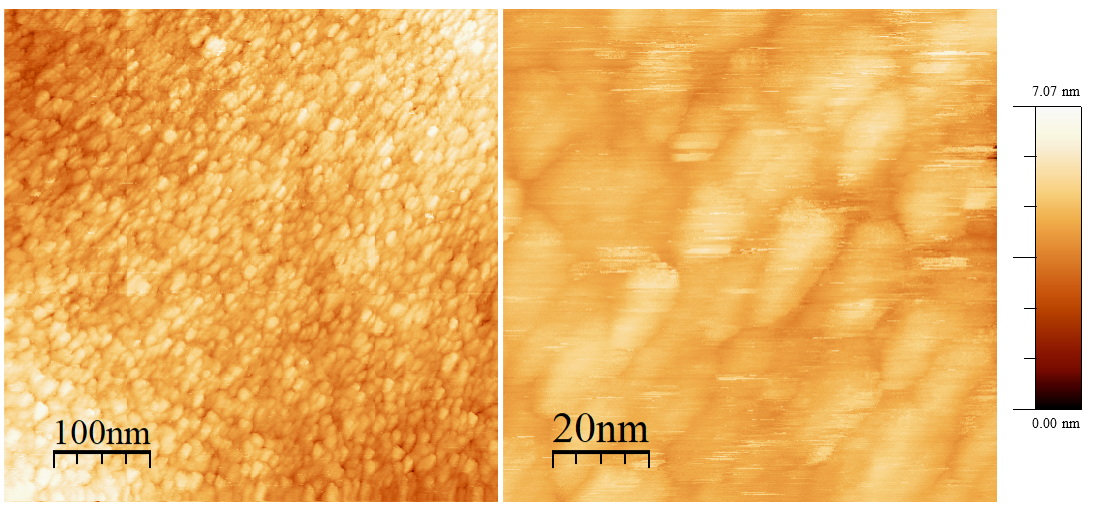}
\caption{Representative STM images of the $t_{\text{Pt}}$= 12 nm sample at two different amplifications.}
\label{STM}
\end{figure}

The surface of the $t_{\text{Pt}}$= 7 nm sample was characterized by Scanning Tunneling Microscopy (STM, Omicron VT SPM) in Ultra High Vacuum (pressure below10$^{-8}$ Torr) at room temperature. Tips were made from Pt/Ir wire, and checked by imaging HOPG surface with atomic resolution.  Images were analyzed with WSxM software \cite{Horcas07} and representative examples are shown in Fig. (\ref{STM}). In there, it can be observed that the film is continuous ant the typical grain lateral size is of the same order than the thickness of the film. We then do not expect significant grain boundary scattering for electrons moving along out-of-plane direction. 

Eventual  interface discontinuities in smaller $t_{\text{Pt}}$ samples could imply that the actual value of $\rho_{\perp}$ differ slightly from bulk Pt resistivity owing to a reduced effective area of conduction. However, even in this scenario, our conclusions would be unaltered since our analysis relies on the \textit{relative} variation of $\rho_{\perp}$ across $T$ and not on its \textit{absolute} value: our observations are explainable if and only if the $\rho$ that determines $\lambda_{\text{s}}$ exhibits \textit{bulk-like} behavior respect to $T$. The later is not modified by a reduced area of conduction, despite the actual value of $\rho_{\perp}$ is.
Interface discontinuities could also affect the areal density of channels available for spin transport and this must be reflected in the value of $g_{\uparrow\downarrow}$. However this is a fitted parameter that does not have influence the other aspects of our analysis.
The evident trends of variation of  $\Delta\alpha_{\text{sat}}$ and $\lambda_{\text{s}}$ with respect to $T$  in Fig 2(a) support by itself our qualitative conclusions, independent of the specific value of $g^{\uparrow \downarrow}$ or $\tau_{\text{s}}$.
\end{appendices}
\section*{Acknowledgments}
This work was supported by FONDECYT Grant No. 3170908, ANID FONDECYT/REGULAR 1201102, and ANID PIA/APOYO AFB180002.
\bibliographystyle{apsrev4-2}
%

\end{document}